\documentclass[11pt]{iopart}
\pdfoutput=1
\expandafter\let\csname equation*\endcsname\relax
  \expandafter\let\csname endequation*\endcsname\relax

\usepackage{citesort}

\usepackage{hyperref}
\usepackage{float}
\usepackage{cite}
\usepackage{subfigure}
\usepackage{url}

\usepackage[pdftex]{graphicx,color}
\usepackage{amsmath,amssymb,amsfonts}
\usepackage{bm}% bold math

\begin{document}

\title{First-Passage Duality}

\author{P. L. Krapivsky}
\address{Department of Physics, 590 Commonwealth
  Avenue, Boston University, Boston, MA 02215, USA}
\author{S. Redner}
\address{Santa Fe Institute, 1399 Hyde Park Road, Santa Fe, New Mexico 87501,
  USA}
\begin{abstract}
  We show that the distribution of times for a diffusing particle to first
  hit an absorber is \emph{independent} of the direction of an external flow
  field, when we condition on the event that the particle reaches the target
  for flow away from the target.  Thus, in one dimension, the average time
  for a particle to travel to an absorber a distance $\ell$ away is
  $\ell/|v|$, \emph{independent} of the sign of $v$.  This duality extends to
  all moments of the hitting time.  In two dimensions, the distribution of
  first-passage times to an absorbing circle in the radial velocity field
  $v(r)=Q/(2\pi r)$ again exhibits duality.  Our approach also gives a new
  perspective on how varying the radial velocity is equivalent to changing
  the spatial dimension, as well as the transition between transience and
  strong transience in diffusion.
\end{abstract}

\section{Introduction}

How long does it take a diffusing particle to travel from a starting point to
a target?  This first-passage, or hitting time, is a fundamental
characteristic of diffusion and has a myriad of applications---to chemical
kinetics~\cite{chandrasekhar1943stochastic,weiss1967first}, options
pricing~\cite{black1973pricing,kou2003first}, and neuronal
dynamics~\cite{gerstein1964random,bulsara1996cooperative,burkitt2006review}
to name a few examples.  An intriguing property of diffusion in spatial
dimensions $d\leq 2$ is that a diffusing particle is certain to reach any
finite-size target, but the average time for this event is infinite.  This
property of eventually reaching any target is known as \emph{recurrence}. The
dichotomy between reaching a target with certainty, but taking an infinite
time for this event to occur for $d\leq 2$, helps make diffusion such a
compelling and vibrant problem even after a century of intensive
study~\cite{polya1921aufgabe,feller1968introduction,spitzer1964,redner2001guide,morters2010brownian}.

In this article, we investigate basic first-passage characteristics of a
simple convection-diffusion system in which a velocity field is either
directed toward or away from a target.  There has been considerable work on
elucidating the characteristics convection-diffusion systems in a variety of
geometries (see,
e.g.,~\cite{koplik1994tracer,redner1996diffusive,redner1997survival,PhysRevE.62.103,choi_margetis_squires_bazant_2005,condamin2007first,mattos2012first,metzler2014first,holcman2015stochastic,grebenkov2016universal}).
Nevertheless, their first-passage properties, even in the simplest settings,
still exhibit surprises, as we present below.

\begin{figure}[ht]
\centerline{\subfigure[]{\includegraphics[width=0.18\textwidth]{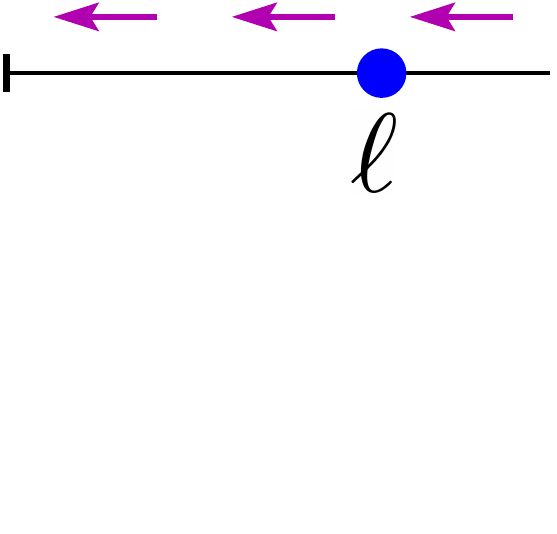}}\qquad\qquad
\subfigure[]{\includegraphics[width=0.3\textwidth]{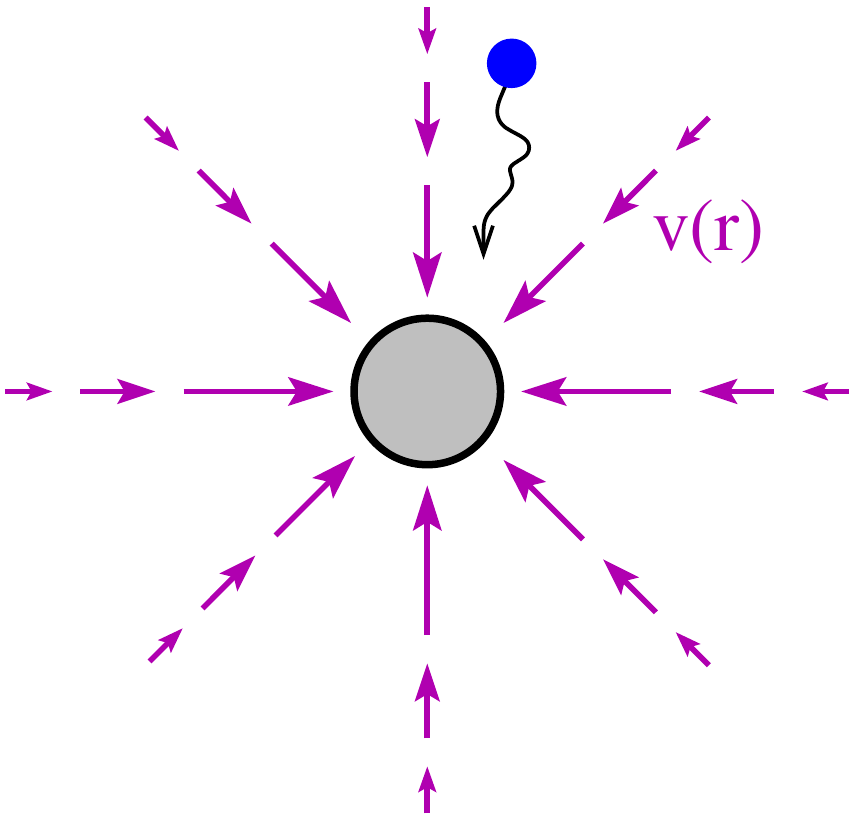}}}
\caption{Illustration of the system in (a) one dimension and (b) two
  dimensions.}
  \label{cartoon}
\end{figure}

Consider first a uniform velocity field of magnitude $|v|$ that drives a
diffusing particle, initially at $\ell>0$, toward a target at $x=0$ in one
dimension (Fig.~\ref{cartoon}(a)).  Since the particle is recurrent in the
absence of convection, the particle will certainly reach $x=0$ when the
velocity is directed toward the target.  The corresponding average hitting
time is finite and equals $\ell/|v|$, as one might naively expect.  Diffusion
plays no role in this average hitting time, but does affect higher moments,
as we will derive below.
  
Conversely, if the velocity is directed away from the target, the probability
for the particle to eventually reach the target is less than 1.  However, for
the fraction of particle trajectories that do reach the target, their average
\emph{conditional} hitting time is again $\ell/v$.  Here, the conditional
hitting time is defined as the average time for those trajectories that
actually reach the target.  For this subset of ``return'' trajectories, they
must reach the target quickly for $v\gg 1$, or else they will be convected
away and never reach the target.  We call the equality between these two
hitting times as \emph{duality}.  As we shall demonstrate, this duality is
quite general and applies for the \emph{distribution} of hitting times
(unconditional for inward flow and conditional for outward flow).

In two dimensions, the natural analog of a constant flow field is radial
potential flow, $\mathbf{v} = {Q\, \hat{\mathbf{r}}}/{2\pi r}$
(Fig.~\ref{cartoon}(b)).  In keeping with this radial symmetry, we define the
absorber to be a circle of radius $a>0$; a non-zero radius is needed so that
a diffusing point particle can actually hit the absorber.  When the flow
field is inward, $Q<0$, a diffusing particle hits the absorber with
certainty, but the average hitting time is finite only for $Q<-4\pi D$, where
$D$ is the diffusion coefficient.  Conversely, for sufficiently strong
outward flow, namely, $Q> 4\pi D$, the conditional average hitting time is
identical to the unconditional average hitting time for inward flow when
$Q<-4\pi D$.  More generally, the distributions of hitting times
(unconditional for inward flow and conditional for outward flow) are
identical.

\section{Duality in One Dimension}
\label{sec:1d}

Let us first treat a diffusing particle with diffusivity $D$ that starts at
$x=\ell>0$ on the infinite line, and is absorbed when $x=0$ is reached. The
particle is also driven by a spatially uniform velocity field of magnitude
$v$.  We first analyze the situation where the drift velocity $v<0$ drives
the particle toward the origin.  All moments of the hitting time can be
computed from the first-passage probability, namely, the probability for the
particle to reach the origin for the first time at time
$t$~\cite{redner2001guide}:
\begin{align}
  F(t) = \frac{\ell}{\sqrt{4\pi Dt^3}}\,\, e^{-(\ell-|v|t)^2/4Dt}\,.
\end{align}
It is straightforward to verify that the eventual hitting probability $H$
equals 1; that is, $H\equiv \int_0^\infty dt\, F(t)=1$.  Since an
isotropically diffusing particle eventually reaches the origin, the origin is
certainly reached when there is a negative drift velocity.

The moments of the hitting time are given by
\begin{subequations}
\begin{align}
  \label{t-def}
 \langle t^n\rangle
  =  \int_0^\infty dt\,\,t^n\frac{\ell}{\sqrt{4\pi Dt^3}}\,\, e^{-(\ell- |v|t)^2/4Dt}\,.
\end{align}
This quantity is properly normalized because the zeroth moment, which
is the eventual hitting probability, equals 1.  To simplify our
results for the hitting time moments, we define the convection time
$t_c\equiv \ell/|v|$, the diffusion time $t_D\equiv \ell^2/(2D)$, and
their ratio $z= t_c/t_D=2D/\ell|v|$; the latter is the inverse of the
P\'eclet number~\cite{probstein2005physicochemical}.  Using these
variables and also introducing the dimensionless time $\tau\equiv t/t_c$,
the moments of the hitting time are
\begin{align}
  \label{tn}
  \langle t^n\rangle & = \int_0^\infty \!\!dt\,t^n \,\frac{\ell}{\sqrt{4\pi Dt^3}}\,\,
  \exp\left[-\frac{\ell^2}{4Dt}+ \frac{|v|\ell}{2D}-\frac{v^2 t}{4D}\right]\,\nonumber\\[3mm]
  &=\frac{t_c^n}{\sqrt{2\pi z}}\,\, e^{1/z}\int_0^\infty \!\!d\tau\,
    \tau^{n-3/2}
    \,\exp\left[-(\tau+\tau^{-1})/2z\right]\,\nonumber\\[3mm]
  &= t_c^n\,\,\sqrt{\frac{2}{\pi z}}\,\, e^{1/z} \,\,K_{\frac{1}{2}-n}(1/z)\,,
\end{align}
\end{subequations}
where $K_\mu$ is the modified Bessel function of the second kind of order
$\mu$~\cite{abramowitz1964handbook}.

From the above expression, the average hitting time is
$\langle t\rangle = \ell/|v|$, in agreement with naive intuition; also notice
that $\langle t\rangle$ is independent of the diffusion coefficient $D$.  The
next few moments are:
\begin{align}
  \langle t^2\rangle& = t_c^2(1+z)\,,\nonumber\\
  \langle t^3\rangle& = t_c^3(1+3z+3z^2)\,,\nonumber\\
  \langle t^4\rangle & = t_c^4(1+6z+15z^2+15z^3)\,,\nonumber
\end{align}
etc.  From these moments, the cumulants~\cite{van1992stochastic} are given
by: $\kappa_1 = t_c$, $\kappa_2 =zt_c^2$, $\kappa_3 = 3z^2t_c^3$, etc.; the
general result is $\kappa_n=(2n-3)!!\,\, z^{n-1}t_c^n$.  The standard
deviation in the hitting time is $\sqrt{\kappa_2}/\kappa_1= \sqrt{t_c/t_D}$.
Thus as the drift velocity $v\to 0$, fluctuations in the hitting time between
different trajectories diverge.  As a curious sidenote, the numerical
coefficient of the $n^{\rm th}$ cumulant is the same as the $n^{\rm th}$
moment of the Gaussian distribution $P(y) = \exp(-y^2/2)/\sqrt{2\pi}$.

For positive drift velocity, $v>0$, the probability that the particle
eventually hits the origin is
\begin{align}
H=  \int_0^\infty dt\, \frac{\ell}{\sqrt{4\pi Dt^3}}\,\, e^{-(\ell+vt)^2/4Dt} =e^{-v\ell/D}\,.
\end{align}
The moments of the hitting time, conditioned on the particle actually
reaching the origin, are given by
\begin{align}
  \label{tn-vm-inf}
\langle t^n\rangle &=\frac{1}{H}  \int_0^\infty dt\,\, t^n \,\frac{\ell}{\sqrt{4\pi Dt^3}}\,\,
  e^{-(\ell+vt)^2/4Dt}\, \nonumber\\
  &= e^{v\ell/D}\,\int_0^\infty dt\,\, t^n \frac{\ell}{\sqrt{4\pi Dt^3}}\,\,
    e^{-(\ell+vt)^2/4Dt}\,\nonumber\\
  &= \int_0^\infty dt\,\,t^n \frac{\ell}{\sqrt{4\pi Dt^3}}\,\,
    e^{-(\ell-vt)^2/4Dt}\,.
\end{align}
When the prefactor $\frac{1}{H}=e^{v\ell/D}$ is combined with the integrand,
its effect is merely to change the sign between $\ell$ and $vt$ in the
exponential in the numerator.  As a result, the last line of
\eqref{tn-vm-inf} is identical to Eq.~\eqref{t-def}.  Thus all moments of the
\emph{conditional} hitting time for convection-diffusion, with drift
\emph{away} from the origin, coincide with the corresponding moments of the
\emph{unconditional} hitting time for convection-diffusion, with drift
\emph{toward} the origin.  This is the statement of first-passage duality in
one dimension.

\section{Duality in Two Dimensions}

We now extend duality to two dimensions.  In two dimensions, the target must
have a non-zero size so there is a positive probability that the target will
be hit by a diffusing particle.  We take the target to be a disk of radius
$a$ and study the hitting time to this disk when the particle starts at an
arbitrary point in the exterior region $r>a$.

The natural two-dimensional analog of a constant velocity field in one
dimension is radial potential flow, with velocity
\begin{equation}
\label{v2d}
\mathbf{v}(\mathbf{r})=\frac{Q}{2\pi r}\,\mathbf{\hat r}\,.
\end{equation}
Such a flow is generated by a point sink (or source) of strength $Q$ at the
origin in an incompressible fluid.  This flow field has the useful property
that in rotationally-symmetric situations the convection term can be absorbed
into the diffusion operator by an appropriate shift of the spatial
dimension~\cite{PhysRevE.62.103,krapivsky2007probing}.  We will exploit this
equivalence between the flow field and the spatial dimension in what follows.
To simplify matters and without loss of generality, we take the initial
condition to be a probability density that is symmetrically concentrated on a
ring of radius $r$.  By this construction, the system is always rotationally
symmetric.

\subsection*{Hitting probability and average hitting time}

In principle, the hitting probability and average hitting time can be
determined by solving the convection-diffusion equation in the flow
field~\eqref{v2d}, then computing the diffusive flux to the absorbing circle,
and finally extracting the eventual hitting probability and the average
hitting time from moments of this flux~\cite{redner2001guide}.  A simpler
approach relies on the backward Kolmogorov
equation~\cite{kolmogorov1931analytic}, which exploits the Markov nature of
the particle motion to write a time-independent equation for the hitting
probability.

For example, for a particle that starts at $r$, the hitting probability to a
target set can be generically written as (for a discrete hopping process for
concreteness)
\begin{align}
  \label{H-gen}
  H(r) = \sum_{r'} p(r\to r')\, H(r')\,,
\end{align}
where $p(r\to r')$ is the probability to hop from $r$ to $r'$ in a single
step.  That is, the hitting probability starting at $r$ may be decomposed
into the sum of hitting probabilities after one step, multiplied by the
probability for the particle to take a single step from $r$ to $r'$.
Expanding \eqref{H-gen} in a Taylor series leads to the backward Kolmogorov
equation for $H(r)$, in which the initial point is the dependent variable.
The nature of this equation depends on the geometry of the system and the
single-step hopping probabilities.

For radial potential flow, the probability $H(r)$ that a particle, which
starts at radius $r$, eventually hits the disk
satisfies~\cite{redner2001guide}
\begin{equation}
\label{Hr:eq}
D\Big(H''+\frac{1}{r}\,H'\Big)+\frac{Q}{2\pi r}\,H' = 0\,,\nonumber
\end{equation}
where prime denotes radial derivative.  We rewrite this equation as
\begin{subequations}
\label{Hab}
  \begin{equation}
\label{Hq}
H''+\frac{1+q}{r}\,H' = 0\,,
\end{equation}
with $q \equiv {Q}/{2\pi D}$ is the dimensionless source strength.  Let us
compare \eqref{Hq} with the corresponding equation
\begin{equation}
\label{Hd}
H''+\frac{d-1}{r}\,H' = 0
\end{equation}
\end{subequations}
for the hitting probability in $d$ dimensions for pure diffusion (without
convection) exterior to the ball.  Equations \eqref{Hq} and \eqref{Hd} are
\emph{identical} when $d=2+q$.  We may thus interpret the hitting probability
for diffusion with potential flow in two dimensions as equivalent to the
hitting probability for isotropic diffusion in spatial dimension $d=2+q$.  By
this correspondence, an increase in $q$ is equivalent to an increase in the
spatial dimension.  This correspondence accords with naive expectations: by
increasing the outward velocity, it becomes less likely that that the
absorber will be reached.  Such a decrease in the hitting probability also
occurs in isotropic diffusion when the spatial dimension is increased.  To
obtain the hitting probability, we solve \eqref{Hq} subject to
$H(r\!=\!a)=1$, and obtain
\begin{equation}
\label{H:2d}
H(r)=
\begin{cases}
1        & \qquad q\leq 0\,,\\
r^{-q}  & \qquad q > 0\,,
\end{cases}
\end{equation}
where we henceforth set $a=1$ and $D=1$ to simplify the formulae that follow.

In a similar spirit to the hitting probability, the average hitting time can
also be obtained by solving an appropriate backward Kolmogorov equation.  In
this case, the underlying equation is
\begin{align}
  \label{t-gen}
  T(r) = \sum_{r'} p(r\to r')\, \big[T(r')+\delta t\big]\,,
\end{align}
where we write $T(r)\equiv\langle t(r)\rangle$ for the average hitting time to the
circle of radius 1 when the particle starts at $r$.  Here, we again decompose
the average hitting time starting from $r$ as the time for a single step (the
factor $\delta t$), plus the average hitting time starting from neighboring
points $r'$, multiplied by the probability of this single step.  Performing
the same Taylor expansion as that used to recast the general equation
\eqref{H-gen} for the hitting probability as \eqref{Hab}, $T(r)$
satisfies~\cite{redner2001guide}
\begin{equation}
\label{Tq}
T''+\frac{1+q}{r}\,T' = -1
\end{equation}
when $q\leq 0$.  We must solve this equation subject to the boundary
condition $T(1)=0$; that is, starting at the disk boundary, the hitting time
is zero.

There exists a one-parameter family of solutions to \eqref{Tq}:
\begin{equation}
  \label{Tq:sol}
  T(r) =
 \begin{cases} {\displaystyle
     \frac{r^2-1}{-2(2+q)}+A\big(r^{-q}-1\big)}&\qquad q\ne -2\,,\\[5mm]
 {\displaystyle -\frac{r^2}{2}\,\ln r+A\big(r^2-1\big)}& \qquad q=-2\,.
\end{cases}
\end{equation}
In the range $-2<q\leq 0$ the solution for $q\ne -2$ is unacceptable, as it
becomes negative for any finite choice of $A$; the solution for $q=-2$ has
the same deficiency.  The resolution of this apparent paradox is simple:
$A=\infty$, so that $T(r)=\infty$ in the range $-2\leq q\leq 0$.  When
$q<-2$, the expression in the first line of \eqref{Tq:sol} with $A=0$ is the
proper solution.  The vanishing of this amplitude follows from the
observation that with an inward convection field $T(r)$ cannot grow
faster than $r^2$ for $r\gg 1$.

When $q>0$, we want to solve for the average \emph{conditional} hitting time.
By generalizing the backward Kolmogorov approach in the appropriate
way~\cite{redner2001guide}, the governing equation for this average
conditional time is
\begin{equation}
\label{Tq:plus}
(HT)''+\frac{1+q}{r}\,(HT)' = -H\,,
\end{equation}
with $H=r^{-q}$ from \eqref{H:2d}.  Solving \eqref{Tq:plus}, subject
$T(1)=0$, gives
\begin{equation}
\label{Tq:sol-plus}
T(r) =
\begin{cases}
 {\displaystyle   \frac{r^2-1}{2(q-2)}+A\big(r^{q}-1\big)} &\qquad q>2\,,\\[5mm]
 {\displaystyle   -\frac{r^2}{2}\,\ln r+A\big(r^2-1\big)}&\qquad  q=2\,.
  \end{cases}
\end{equation}
The same argument as that used for the inward flow field shows that there
are no acceptable solutions in the range $0\leq q\leq 2$; instead $T=\infty$.
When $q>2$, the average hitting time is given by the first line of
\eqref{Tq:sol-plus} with $A$ set to 0.

To summarize, the average hitting time to a disk of radius $a$ is
\begin{equation}
\label{Tq:all}
T(r) = 
\begin{cases}
 {\displaystyle \frac{r^2-a^2}{2(|q|-2)D}} & \qquad |q|>2\\[4mm]
 {\displaystyle \infty                   }         & \qquad |q|\leq  2\,.
\end{cases}
\end{equation}
where the physical variables $a$ and $D$ have been restored.  Crucially,
$T(r)$ depends only on $|q|$.  This is the statement of duality in two
dimensions; namely the average hitting time is \emph{independent} of the sign
of the velocity.  Another important feature is that the average hitting time
(either conditional or unconditional) is finite only for $|q|>2$.

Equation~\eqref{Tq:all} can be given an additional meaning by exploiting the
aforementioned correspondence $d=2+q$ between magnitude of the flow field in
convection diffusion and the spatial dimension in isotropic diffusion.  For
isotropic diffusion, it is known that spatial dimension $d=4$ demarcates the
transition between transience and strong
transience~\cite{jain1968range,peterson2015}.  Transience is the familiar
property that a diffusing particle may not necessarily hit a finite absorbing
set, a property that occurs when the spatial dimension $d>2$.  However, for
the subset of diffusing trajectories that do reach the absorber, the average
conditional hitting time is infinite for $2<d\leq 4$.  However, for spatial
dimension $d>4$, this average conditional hitting time becomes finite, a
features that is known as strong transience.  Thus by varying the strength of the
flow, one can vary the effective dimensionality and drive the system from
recurrent, to transient, to strongly transient.

When $q$ passes through $-2$, the average unconditional hitting time changes
from infinite to finite, while the eventual hitting probability always equals
1.  We term this change as a transition between recurrence (hitting
probability equals 1 and infinite average hitting time) and \emph{strong
  recurrence} (hitting probability equals 1 and finite average hitting time).
By the connection $d=2+q$ between convection-diffusion in two dimensions and
isotropic diffusion in spatial dimension $d$, this transition between
recurrence and strong recurrence for pure diffusion occurs when the spatial
dimension $d=0$.

\subsection*{Distribution of the hitting time}

Duality can be extended to the full distribution of hitting times.  Let
$F(r,t)$ denote the first-passage probability, namely, the probability that
the particle first hits the disk at time $t$ when starting at $r$.  While the
moments of hitting time can be computed by the backward Kolmogorov equation
approach, a more efficient strategy is to write the backward equation for the
Laplace transform of the distribution of hitting times
\begin{equation}
\label{Lap}
\Pi(r,s) = \int_0^\infty dt\,e^{-st}\,F(r,t) = \left\langle e^{-st}\right\rangle\,,
\end{equation}
from which all moments (and cumulants) can be extracted.

Following this approach, the function $\Pi(r,s)$
satisfies~\cite{krapivsky2010kinetic}
\begin{equation}
\label{Pi}
\Pi''+\frac{1+q}{r}\,\Pi' = s\Pi\qquad  q\leq 0\,,
\end{equation}
subject to the boundary condition $\Pi(1,s)=1$, as the hitting time
vanishes at $r=1$.  The general solution to \eqref{Pi} can be expressed in
terms of the modified Bessel functions~\cite{abramowitz1964handbook}:
\begin{equation*}
%\label{Psr}
\Pi(r,s) = r^{-\lambda}\left[A_1 I_{-\lambda}\big(r\sqrt{s}\big) + A_2 K_{-\lambda}\big(r\sqrt{s}\big)\right]\,,
\end{equation*}
where $\lambda=q/2$.  We fix the constants by the boundary conditions.  The
Laplace transform obeys the obvious bounds $0<\Pi(r,s)<1$, with the upper
bound arising from $\Pi(r,s)<\Pi(r,0)=1$.  However, the above general
solution diverges when $r\to\infty$.  To ensure that the solution remains
finite in this limit, we must choose $A_1=0$.  The amplitude $A_2$ is fixed
by the boundary condition $\Pi(1,s)=1$, from which the Laplace transform of
the hitting time distribution is
\begin{subequations}
\begin{equation}
\label{Pi:minus}
\Pi(r,s) = r^{-\lambda}\,\frac{K_{-\lambda}\big(r\sqrt{s}\big)}{K_{-\lambda}\big(\sqrt{s}\big)}
\qquad q\leq 0\,.
\end{equation}
The same result was derived in~\cite{redner2001guide} by directly solving the
convection-diffusion equation for pure diffusion in $d$ dimensions and then
computing the diffusive flux to the absorbing ball.  To make the
correspondence with \eqref{Pi:minus}, we need to replace $d$
in~\cite{redner2001guide} by $2+q$, as discussed above.

When $q>0$, we need to work with the conditional hitting probability.  Thus
the relevant quantity is the function $\Xi(r,s)\equiv H(r)\Pi(r,s)$, which
satisfies the backward equation
\begin{equation*}
%\label{F:eq}
\Xi''+\frac{1+q}{r}\,\,\Xi' = s\,\Xi\,,
\end{equation*}
which is mathematically identical to \eqref{Pi}. Hence its solution, subject
to the boundary condition $\Pi(1,s)=1$, is identical to \eqref{Pi:minus}.  We
now use $H=r^{-q}=r^{-2\lambda}$ and the identity
$K_\lambda(z)=K_{-\lambda}(z)$ to ultimately find
\begin{equation}
\label{Pi:plus}
\Pi(r,s) = \frac{\Xi(r,s)}{H(r)}
=r^{\lambda}\,\frac{K_\lambda\big(r\sqrt{s}\big)}{K_{\lambda}\big(\sqrt{s}\big)}\qquad q>0
\,.
\end{equation}
\end{subequations}
Equations \eqref{Pi:minus} and \eqref{Pi:plus} exhibit the fundamental
duality in the Laplace transform of the first-passage probability with
respect to the transformation $q\longleftrightarrow -q$:
\begin{equation}
  \label{duality}
\Pi(r,s; -q) = \Pi(r,s; q)\,.
\end{equation}

From the series representation of $\Pi(r,s)$, we can extract all moments of
the hitting time and show that there is a transition in the average hitting
time when $q=\pm 2$, as already found in \eqref{Tq:all}, as well as a series
of transitions for progressively higher moments for $q=\pm 4, \pm 6,\ldots$.
Consider the asymptotic behavior of $\Pi(r,s)$ in \eqref{Pi:plus} as
$s\to 0^+$.  For $\lambda>0$ and non integer, we use the identity
\begin{equation}
\label{KI}
K_\lambda(z) = \pi\,\frac{I_{-\lambda}(z)-I_\lambda(z)}{2\sin(\pi\lambda)}\,,
\end{equation}
and the Frobenius series for $I_{\lambda}$~\cite{abramowitz1964handbook}
\begin{equation}
\label{I:series}
I_\lambda(z) = \sum_{n\geq 0}\frac{\left({z}/{2}\right)^{2n+\lambda}}{\Gamma(n+1)\,\Gamma(n+\lambda+1)}
\end{equation}
to find, as $s\to 0^+$,
\begin{equation}
\label{Psr:small}
\Pi(r,s) = 1 + \left(\frac{s}{4}\right)^\lambda \frac{\Gamma(1-\lambda)}{\Gamma(1+\lambda)}\,\big(1-r^{2\lambda}\big) + O(s)\,,
\end{equation}
when $0<\lambda<1$.  If the average hitting time was finite, the Laplace
transform would have the Taylor-series expansion
\begin{equation*}
\Pi(r,s) = 1 - s\langle t(r)\rangle + \ldots\,.
\end{equation*}

Comparing with \eqref{Psr:small} we see that $\langle t\rangle=\infty$ when
$0<\lambda<1$. When $\lambda=1$ we use the known asymptotic behavior of the
Bessel function $K_1(z)$~\cite{abramowitz1964handbook} to obtain
$\Pi(r,s) - 1 \sim s\ln s$.  The absence of a linear term in the expansion
again shows that $\langle t\rangle=\infty$ when $\lambda=1$.  When
$1<\lambda<2$ we can use again \eqref{KI}--\eqref{I:series} to obtain
\begin{equation}
\Pi(r,s) = 1 - s\langle t(r)\rangle + O\big(s^{\lambda}\big)\,,
\end{equation}
with $\langle t\rangle$ given by \eqref{Tq:all}.  However, the second
moment $\langle t^2\rangle$ still diverges when $1<\lambda\leq 2$.  By
using \eqref{KI}--\eqref{I:series} and focusing on the range $2<\lambda<3$,
we find
\begin{equation}
\Pi(r,s) = 1 - s\langle t(r)\rangle + \tfrac{1}{2}s^2 \langle t(r)^2\rangle + O\big(s^{\lambda}\big)
\end{equation}
Thus the first two moments are finite, while the third moment diverges when
$2<\lambda\leq 3$.  The second moment is, explicitly,
\begin{equation}
\label{tav-2}
\langle t(r)^2\rangle =
\begin{cases}
{\displaystyle \frac{r^4 - a^4}{4(|q|-2)(|q|-4)D^2} - \frac{(r^2-a^2)a^2}{2(|q|-2)^2
  D^2}}&\qquad |q|>4\,,\\[4mm]
\infty  & \qquad |q|\leq 4\,,
\end{cases}
\end{equation}
while the variance is described more compactly as
\begin{equation}
\label{var}
\langle t^2\rangle - \langle t\rangle^2 = \frac{r^4 - a^4}{2(|q|-2)^2 (|q|-4) D^2}
\end{equation}
when $|q|>4$.
The transition from the second moment being infinite to being finite when $q$
passes through 4 corresponds to a transience/strong transience transition for
the second moment for isotropic diffusion in spatial dimension $d=6$, while
the transition when $q$ passes through $-4$ corresponds to a
recurrence/strong recurrence transition in the second moment for isotropic
diffusion when $d=-2$.  Generally, when $n<\lambda\leq n+1$, the moments
$\langle t^j\rangle$ exist when $j=0,1,\ldots,n$ and diverge when
$j\geq n+1$.

\section{Discussion}

We discovered a simple duality for the distribution of hitting times to an
absorber for a diffusing particle in a constant ($1d$) or radial potential
($2d$) velocity field.

In one dimension with flow toward the absorber, all particle trajectories
eventually hit the target in a finite time and the basic quantity is the
\emph{unconditional} hitting time and all its moments.  When the flow is away
from the absorber, it is necessary to restrict to the \emph{conditional}
hitting time, defined as hitting time of the subset of trajectories that
eventually reach the target (and all its moments).  We showed that all
moments of the unconditional hitting time for flow toward an absorber
coincide with the corresponding moments of the conditional hitting time for
flow away from the absorber.  As a consequence, the conditional hitting time
\emph{decreases} when the outward flow velocity \emph{increases}.

Our derivations in one dimension relied on direct calculations, while in two
dimensions we employed the backward Kolmogorov equation.  This latter
approach can be also used in one dimension.  Within this approach, there is a
subtlety in the determination of the average hitting time for inward flow.
Mathematically, one must solve a second-order equation
\begin{align}
  \label{T:disc}
  D\frac{d^2 T}{d\ell^2}-v\frac{d T}{d\ell}=-1\,,
\end{align}
subject to a single boundary condition T(0) = 0. The one-parameter family of
solutions to  \eqref{T:disc} that satisfies this boundary condition is
\begin{align}
  \label{T-last-eq}
  T(\ell) =\frac{\ell}{v} + A \big(e^{v\ell/D}-1\big)\,.
\end{align}
Setting the amplitude $A=0$, gives the correct (and unique) solution.  The
simplest way to show that $A=0$ is by physical reasoning.  For $v\to\infty$,
the second term in \eqref{T-last-eq} diverges, whereas the hitting time must go to
zero.  Thus $A=0$.  One can also obtain $T(\ell)$ by solving this problem in
the finite interval $[0,L]$ with absorption at $0$ and reflection at $L$, and
then take the limit $L\to\infty$.  By either approach, the average hitting
time is simply $T(\ell)=\ell/v$.

The duality between hitting times also holds in the interval $[0,L]$ with
both boundaries absorbing.  Here, the basic observables are the hitting
probabilities and the conditional hitting times to $0$ and to $L$.  The
approach of Sec.~\ref{sec:1d} now gives the following: for positive drift
velocity $v>0$, the probability to hit the origin is
\begin{subequations}
\begin{equation}
  \label{H-vp-int}
  H(\ell)= %1- \frac{(1-e^{-v\ell/D})}{(1-e^{-vL/D})} =
  \frac{e^{-v\ell/D}-e^{-vL/D}}{(1-e^{-vL/D})}\,,
\end{equation}
while the average conditional hitting time to the origin is
\begin{align}
  \label{t-vp-int}
  T(\ell) &= \frac{\ell}{v}\,\,\frac{e^{-vL/D}+e^{-v\ell/D}}{e^{-v\ell/D}-e^{-vL/D}} -\frac{2L}{v} \,\,\frac{1-e^{-v\ell/D}}{1-e^{-vL/D}} \,\,\frac{e^{-vL/D}}{e^{-v\ell/D}- e^{-vL/D}}\,.
\end{align}
\end{subequations}
For $L\to\infty$, we recover the result of the semi-infinite system,
$H(\ell)=1$ and $T(\ell)=\ell/v$.  When $v<0$, the hitting probability to the
origin is again given by \eqref{H-vp-int}, but with the opposite sign for
$v$.  This expression is not invariant under the interchange $v\to -v$.
However, the average conditional hitting time \eqref{t-vp-int} \emph{is}
invariant under the interchange $v\to -v$.  A similar duality for the finite
interval was found previously in
Refs.~\cite{PhysRevLett.115.250602,PhysRevX.7.011019} and was extended to
general potentials (not just constant drift) in
Refs.~\cite{PhysRevLett.97.020601,doi:10.1063/1.3179679}.

In two dimensions, we showed that the distribution of unconditional hitting
times for inward flow is the same as the distribution of conditional hitting
times for outward flow, for the radial potential velocity field
$v=Q/(2\pi r)$.  Consequently, all moments of the unconditional hitting times
for inward flow coincide with the corresponding moments of the conditional
hitting times for outward flow.  This relationship between hitting times
follows from the connection between convection-diffusion in two dimensions in
the presence of a radial flow field and pure diffusion in general spatial
dimensions that was previously studied in Ref.~\cite{krapivsky2007probing}.
In this work, we exploited this connection to derive a duality in the
first-passage properties of radial flow.

A basic feature in two dimensions is that the average hitting time
(unconditional for inward flow, conditional for outward flow) diverges for
sufficiently weak flow, namely, $|Q|\leq 4\pi D$.  We also showed that
increasing the magnitude of the outward flow is equivalent to increasing the
spatial dimension $d$ through the relation $d=2+q$, with $q=Q/(2\pi D)$.
Thus the transition between finite and infinite conditional hitting time as
$q$ passes through 2 also describes the transition between transience and
strong transience~\cite{jain1968range,peterson2015} as the spatial dimension
of a system with isotropic diffusion passes through 4.

The duality between $q\leftrightarrow -q$ also has an intriguing consequence
for isotropic diffusion.  Because of the equivalence between potential flow
in two dimensions with scaled velocity $q$ and isotropic diffusion in spatial
dimension $d=2+q$, the duality $q\leftrightarrow -q$ translates to the
duality $d \leftrightarrow 4-d$ for isotropic diffusion.

Letting the spatial dimension $d$ to be a free parameter and then developing
theoretical approaches based on this parameter, such as expansion about a
critical dimension, dimensional regularization, etc., has proven insightful
in field theory and statistical
physics~\cite{wilson1972critical,wilson1972feynman,itzykson1989statistical}.
Such non-integer dimensions arise naturally in many theories of critical
phenomena as an intermediate step toward understanding physically relevant
spatial dimensions, such as $d=2$ and $d= 3$.  We demonstrated a connection
between: (a) two-dimensional convection-diffusion with a radial potential
flow field of scaled magnitude $q$, and (b) isotropic diffusion in spatial
dimension $d=2+q$.  It would be exciting to realize these
convection-diffusion flows experimentally and thereby probe dynamical
processes in spaces whose spatial dimension is not necessarily an integer.

PLK acknowledges the hospitality of the Santa Fe Institute for support of a
research visit to the SFI.  SR acknowledges financial support from grant
DMR16-08211 from the National Science Foundation.

\newpage
\subsection*{References}

%\bibliography{duality}

\providecommand{\newblock}{}

\end{document}